\documentstyle[a4,12pt]{article}
\title{\ \\[-7ex] \hrulefill \\ {\bf Renormalisation-theoretic 
analysis of non-equilibrium phase transitions I:  \\ The \BD\ equations 
with \\ power law rate coefficients} \\[-1ex] }
\author{Jonathan AD Wattis$\dag$ and 
Peter V Coveney$\ddag$\\[-0.5ex] 
{\footnotesize $\dag$Division of Theoretical Mechanics, 
School of Mathematical Sciences,}\\[-0.5ex]
{\footnotesize University of Nottingham,  University Park, Nottingham, 
NG7 2RD, U.K.}\\[-0.5ex]
{\footnotesize $\ddag$Centre for Computational Science, 
Department of Chemistry,}\\[-0.5ex]
{\footnotesize Queen Mary, 
University of London, Mile End Road, London, E1 4NS.}\\[-0.5ex]
{\footnotesize$\dag$ \verb$Jonathan.Wattis@nottingham.ac.uk$ 
\hspace*{8mm}$\ddag$\verb$p.v.coveney@qmul.ac.uk$}\\[-2ex]}
\date{{\footnotesize 31$^{{\rm th}}$ July, 2001} 
\\[-1ex] \hrulefill}

\newcommand{\beqa}{\begin{eqnarray}}
\newcommand{\eeqa}{\end{eqnarray}}
\newcommand{\beq}{\begin{equation}}
\newcommand{\eeq}{\end{equation}}
\newcommand{\rec}[1]{\mbox{$\frac{1}{#1}$}}
\newcommand{\mfrac}[2]{\mbox{$\frac{#1}{#2}$}}
\newcommand{\pad}[2]{\frac{\partial #1}{\partial #2}}
\newcommand{\padd}[2]{\frac{\partial^2 #1}{\partial #2^2}}
\newcommand{\half}{\mbox{$\frac{1}{2}$}}

\newcommand{\nn}{\nonumber}
\newcommand{\ds}{\displaystyle}
\newcommand{\BD}{Becker-D\"{o}ring}
\newcommand{\etal}{{\em et al.}}
\newcommand{\ro}{\varrho}
\let\tilde=\widetilde
\let\hat=\widehat
\newcommand{\rsum}{\sum_{r=1}^{\infty}}
\newcommand{\nsum}{\sum_{n=1}^{\infty}}
\newcommand{\psum}{\sum_{p=1}^{\infty}}
\newcommand{\ksum}{\sum_{k=1}^{\infty}}
\newcommand{\ep}{\varepsilon}

\newcommand{\erfc}{{\rm erfc}}

\newcommand{\ol}[1]{\overline{#1}}
\newcommand{\lbl}[1]{\label{#1}}
\newcommand{\slbl}[1]{\label{#1}}
\begin{document}
\renewcommand{\theequation}
{\arabic{section}.\arabic{equation}}
\newfont{\bbold}{msbm10 scaled\magstep1}
\newcommand{\bbf}[1]{\mbox{{\bbold #1}}}

%----------------------------------------

\maketitle

\vspace*{-15mm}

\begin{abstract}
\noindent We study in detail the application of renormalisation theory to 
models of cluster aggregation and fragmentation of
relevance to nucleation and growth processes.  We investigate 
the \BD\ equations, originally 
formulated to describe and analyse non-equilibrium phase 
transitions, and more recently generalised to 
describe a wide range of physicochemical problems. 
In the present paper we analyse how the systematic
coarse-graining renormalisation of the \BD\ system of equations affects 
the aggregation and fragmentation rate coefficients.
We consider the case of power-law size-dependent 
cluster rate coefficients which we show lead to only three 
classes of system that require analysis:
coagulation-dominated systems, fragmentation-dominated 
systems and those where 
coagulation and fragmentation are exactly balanced. 
We analyse the late-time asymptotics associated with 
each class.
\end{abstract} 
\vspace*{-3mm}
\hrulefill

\noindent {\bf PACS}: \\
64.60.-i General studies of phase transitions \\ 
64.60.Ht Dynamic critical phenomena \\ 
82.20.-w Chemical kinetics

\noindent {\bf Keywords}: \\ 
Nonequilibrium phenomena\\ 
Renormalisation group

%\newpage
%\tableofcontents
%\normalsize
%\noindent
%\vspace*{-5mm}
\normalsize
\newpage
%------------------------------------------------

\section{Introduction} \setcounter{equation}{0}

The purpose of this paper is to develop renormalisation
methods for nucleation and growth processes. 
Renormalisation theory has been applied to various 
problems in theoretical physics, being widely used 
in high energy physics~\cite{itzyk80} and in the analysis 
of equilibrium phase transitions in statistical physics~\cite{cardy}.   
It has also been used in a limited way in asymptotic analysis 
by Goldenfeld~\cite{goldbook}. However, it has not 
previously been used in the asymptotic analysis of 
non-equilibrium phase transitions of the kind
to be discussed in the present paper, mainly 
due to the complex nature of these models which 
often involve nonlinear and non-local features.

In statistical mechanics, the basic idea underlying the 
concept of renormalisation is the systematic filtering out 
of unwanted degrees of freedom in the transition from 
the microscopic to the macroscopic description of 
phenomena. In the case of equilibrium phase transitions, 
near a critical point the system looks the same on all 
length scales and this physical insight is translated into 
mathematical language in terms of `the renormalisation 
group' (RG), a set of scale-invariant transformations which
leave the essential physical properties of the system 
unchanged. Admittedly, the RG terminology is somewhat inappropriate 
since the symmetry of such transformations is that of a 
semi-group (information is lost on coarse-graining), 
and the methodology comprises many different ideas and
distinct methods, rather than being a formal monolithic 
edifice. Wilson's work on the renormalisation group spawned 
a vast effort in the  analysis of equilibrium critical phenomena 
and quantum field theory~\cite{wilson,amit}. Of much more recent 
interest has been the application of the renormalisation 
methodology to non-equilibrium phenomena. The 
complexity of far from equilibrium dynamics is such
that, while the physical motivation behind the RG 
programme of coarse-graining microscopic models 
still seems appropriate to obtain the macroscopic properties,
it must be shown in each specific case that there may be a
suitable scale-invariance (``renormalisation group'') underlying 
the physics, which can reveal  or explain universality in behaviour.

Of central interest in the present paper is the asymptotic late time
macroscopic behaviour of complex dynamical systems.
Previously, Bricmont \& Kupiainen have combined 
asymptotic methods with ideas from renormalisation 
theory to study problems in diffusive processes and 
special limiting solutions of nonlinear parabolic
equations~\cite{bk91,bk92,bk94} while 
Woodruff has cast multiple timescale problems in the form 
of renormalisation theory~ \cite{w93,wm94,w95,w95b}. 
This allowed the equations governing larger-scale 
phenomena to be derived from a more general theory 
and to be separated out from the small scale dynamics. In his 
approach, Woodruff separates timescales in the asymptotic limit 
$\ep\rightarrow0$ by writing the time $t$ as $t = 
\ol{t}/\ep +\tilde{t}$ where $\tilde{t}$ is the fast timescale 
and $\ol{t}$ is the slow timescale. The methods we develop
in this paper share some of these features.

We apply underlying concepts from renormalisation 
theory to study a simple model of nucleation, namely the 
\BD\ equations.  These were originally formulated to 
study first-order phase transitions, and have been a central pillar in
classical nucleation theory. They describe the stepwise growth and
fragmentation of clusters in terms of the rates of the individual
processes wherein monomer particles join or leave each cluster. 
Despite the age of this model, we have only very recently 
applied generalisations of the \BD\ equations 
to a wide range of physicochemical processes from those involving 
surfactant self-assembly~\cite{cw96,cw98} through
RNA-polymer-formation~\cite{ch90,ch95,wc98} 
to cement-setting~\cite{ch95,wc97}. 
In these studies coarse-graining procedures are derived and 
used to reduce large systems of equations down to 
progressively lower-dimensional--``mesoscopic''--dynamical 
systems capable 
of theoretical analysis using standard techniques 
{}from the modern theory of differential equations.  The coarse-grained 
contraction procedure, which is summarised in Section 2,
is analogous to other renormalisation methods used in 
statistical physics, 
and this feature is exploited within this paper.

The \BD\ equations have recently been subjected to 
more conventional analysis using matched 
asymptotic expansions~\cite{wk98,wk99}. 
Rather than use the coarse-graining approximation, which 
emphasises the discrete nature of the equations, that 
analysis concentrated on the large-time 
limit where continuum approximations become valid~\cite{wk98}.  
The key results from this analysis are summarised 
here in Section 3. 
The large-time limit of the coarse-grained equations was also 
studied there, and revealed the necessity of combining a change 
of timescale with the coarse-graining process in order to 
preserve the accuracy of the dynamics of the system; in the present
analysis, this translates into the need to use a dynamical
renormalisation scheme.

Asymptotic methods have been applied directly to 
coagulation-fragmentation problems before, for example 
van Dongen \& Ernst \cite{dongen+e,dongen+e2} and 
Davies \etal\ \cite{dkw98} have  elucidated the self-similarity 
of the pure aggregation form of Smoluchowski's equations. 
These equations differ from the \BD\ equations in 
allowing clusters arbitrary size to coalesce together 
whereas the \BD\ equations only allow cluster-monomer 
interactions, but allow both aggregation and fragmentation. 
The Smoluchowski coagulation-fragmentation problem 
is much more complex but has been analysed by van 
Dongen \& Ernst \cite{dongen+e3} and Carr \& da Costa 
\cite{carr, cc,fpc}. 
The scaling behaviour of the \BD\ equations 
has been studied by several authors, for example, 
Brilliantov \& Krapivsky \cite{bk} and 
Blackman \& Marshall \cite{b+m}. 
Coarse-grained versions 
of the \BD\ equations have been analysed using matched 
asymptotic methods~\cite{compBD},~\cite{wc99}.  
In addition many authors have considered hybrid 
systems which have combined features of the 
Smoluchowski equations and the \BD\ equations: 
this approach started with Samsel \& Perelson \cite{sp} 
who studied a system in which two clusters of arbitrary 
size can coalesce, but with a fragmentation step 
which is strictly \BD\ in nature, allowing only monomers 
to dissociate from clusters.   Such models have been 
investigated further by Krapivsky \& Redner \cite{krap+r} 
and Hendricks \& Ernst \cite{hend}.  Krapivsky \cite{krap} 
considered an alternative modification of the \BD\ equations, 
again concentrating analysis on the large-time scaling laws, while 
Peliti \cite{peliti} looked at the scaling behaviour of 
a very simple model of pure aggregation in various 
dimensions in order to discover the manner in which the 
scaling laws depend on spatial dimension.

Our methods fit within the spirit of Woodruff's approach if we
write the microscopic 
aggregation number $r$ of a cluster as $r = (\lambda-1)n+1+k$ 
where $n$ is of mesoscopic size and $k$ is a microscopic 
correction; we then aim to determine the problem on 
the mesoscale in a form  which does not require us to 
simultaneously solve the microscopic problem.  Thus
microscopic detail is filtered out, but we are able to 
construct a simpler 
model which remains valid on larger (mesoscopic and 
macroscopic) scales. 
Velazquez has recently used a 
renormalisation technique in an attempt to draw 
together the theories of Lifshitz-Slyozov coarsening 
and nucleation as modelled by the \BD\ 
equations~ \cite{velaz}. 
His renormalisation group differs from
ours in several ways, most egregiously in that it separates 
the monomer concentration 
{}from the large-cluster sizes, whereas we retain the 
monomer concentration as an integral part of the 
mesoscale model.

Preliminary results of our work were reported in an earlier 
publication~\cite{rg}.   Here we provide the details 
omitted from that communication. We also discuss how 
the coarse-graining procedure affects the large-time 
kinetic behaviour of the systems under consideration.  
Various types of system are introduced and analysed, 
including full and coarse-grained systems of \BD\ equations 
(Sections~\ref{BDs} and~\ref{coarse}), the underlying 
renormalisation structure described (Section~\ref{renorm}), 
and the large time asymptotics associated with the fixed 
points of the renormalisation mapping are
analysed (Section~\ref{asymptotics}). We consider 
noise-free \BD\ systems, by which we mean those with 
simple analytical formulae for their rate 
coefficients; in a separate paper \cite{rgpap2} we present 
our analysis of  ``noisy'' systems in which the rate constants 
are perturbed by a random amount from the analytical 
expressions used here.  We conclude with a discussion of our 
results (Section~\ref{discussion}).

%-----------------------------
\section{The \BD\ cluster equations} 
\lbl{BDs} \setcounter{equation}{0}
%\subsection{The \BD\ cluster equations} 

In this section, we give a basic outline 
of the \BD\ system of equations and their properties.  
We consider the case for which the monomer concentration
($c_1$) is held constant; thus the \BD\ equations we 
are concerned with here are
\beq
\dot c_r =  J_{r-1} - J_r,\;\;\; (r\geq2), \;\;\;\;\;\;
J_r = a_r c_r c_1 - b_{r+1} c_{r+1} ,\lbl{fullBD} 
\eeq 
and we leave the constant mass formulation (in which 
the monomer concentration may vary)
for future study.   In this system of 
equations, $c_r(t)$ represents the concentration of 
clusters containing $r$ monomers.  Clusters of any 
particular size can be formed by two processes: 
either by the next smallest cluster size coalescing 
with a monomer, or by the next largest size losing 
a monomer.  Only such monomer--cluster interactions 
are permitted in the \BD\ model of nucleation; 
cluster--cluster interactions are ignored.    The 
assumption of a constant monomer concentration 
is made in situations where the so-called ``pool chemical 
approximation'' is valid, namely where there is a large 
source of monomer species entering into the system at 
a rate which maintains the monomer 
concentration essentially fixed and independent of time.

There are certain special properties of the \BD\ 
system which are not immediately apparent from the 
equations~(\ref{fullBD}), and which must be preserved under 
the coarse-graining rescaling if the method is to produce 
physically relevant results, through approximating the 
original system in a faithful manner. These are:

\begin{description}
\item[(i)] There exists a partition function, $Q_r$, satisfying 
$a_rQ_r=b_{r+1}Q_{r+1}$, and hence an equilibrium 
solution $c_r=Q_rc_1^r$.  The quantity $Q_r$ is 
related to the chemical potential of a cluster of 
size $r$. If we denote Boltzmann's constant by $k$, 
the temperature of the system by $T$ and use the 
reference state $\mu_r^\ominus$ in which 
$\mu_1^\ominus=0$, then the chemical potential 
of a cluster of size $r$ (denoted $\mu_r$) is given by 
$\mu_r = \mu_r^\ominus + k T \log c_r$. At 
equilibrium, this must satisfy $\mu_r=r\mu_1$,  
thus $\mu_r^\ominus=-kT\log Q_r$.
\item[(ii)] There exists a Lyapunov function, $V = \rsum c_r 
( \log(c_r/Q_rc_1^r) - 1)$,  
which is related to the Helmholtz free energy of the system by 
$F=k T{\cal V}V$, where ${\cal V}$ is the volume of the system.
\item[(iii)] There exists a set of identities (or ``weak form''), 
$\ds\sum_{r=2}^\infty g_r \dot c_r = g_1 J_1 + 
\ds \rsum \,[\, g_{r+1} - g_r \,]\, J_r$. 
\item[(iv)] The total density of the system is $\ro = 
\ds\rsum r c_r$, however this is not conserved 
since matter can be added to or removed from 
the system by he monomer concentration being 
held constant. 
\end{description}

For certain choices of rate coefficients $a_r,b_r$ and 
certain monomer concentrations $c_1$, the 
equilibrium solution will not decay to zero in the 
limit $r\rightarrow\infty$. In these cases, an alternative 
steady-state solution will be approached in the large-time 
limit. This solution is given by a constant nonzero flux through 
the system, that is $J_r=J$ independent of $r$. This 
condition yields the family of solutions 
\beq
c_r = Q_r c_1^r \left( 1 - J \sum_{k=1}^{r-1} 
\frac{1}{a_k Q_k c_1^{k+1}} \right) , \lbl{gsss}
\eeq
which contains the equilibrium solution as the 
special case $J=0$. 
The steady-state flux $J$ is determined by 
requiring the concentrations $c_r$ to decay to 
zero in the large-$r$ limit, giving 
\beq
J=1\left/\sum_{r=1}^{\infty}\frac{1}{a_rQ_rc_1^{r+1}}\right. .
\eeq

%--------------------------
\subsection{Other formulations}

An alternative formulation of the \BD\ equations 
has been proposed by Penrose \& Lebowitz \cite{pl76}, 
in which the monomer concentration is allowed to 
vary, whilst the total mass of material in the system ($\ro $) 
is held constant.  The analysis of such a system is in 
general more complicated since the monomer 
concentration is unknown and causes the system 
to become both nonlinear and non-local (in $r$).

An underlying model in the present work is a 
constant mass system in which monomer is added to 
or removed from the system by way of a precursor 
chemical $P$ which undergoes a reversible reaction 
to create monomer $C_1$.  This is modelled by the 
simple reaction $P \rightleftharpoons C_1$ with 
forward rate $k_f(p,c_1)$ and reverse rate $k_b(p,c_1)$.  
Denoting the concentration of $P$ by $p(t)$ we then 
have the system 
\beqa
\dot p & = & k_b(p,c_1) c_1 - k_f(p,c_1) p , \\ 
\dot c_1 & = & k_f(p,c_1) p - k_b(p,c_1) c_1 - J_1 - 
\sum_{r=1}^\infty J_r , 
\eeqa
in addition to (\ref{fullBD}). This system has a 
conserved quantity $\ro_{{\rm tot}} = p + \sum_{r=1}^\infty 
r c_r$, however note that the amount of mass in the \BD\ 
part of the system ($\{c_r\}_{r=1}^\infty$) is not constant 
$\ro_{{\rm BD}}(t)=\sum_{r=1}^\infty r c_r$.  Such an extended 
system of the \BD\ equations has been used previously to 
model generalised nucleation phenomena \cite{wc97}. 
In certain cases analytical progress in the study of such 
systems can be made using asymptotic techniques, 
such as in \cite{wk98} where the constant mass formulation 
of the \BD\ equations are investigated, and in \cite{wc97} 
where a generalised constant mass system with inhibition 
as well as a precursor chemical is analysed.

In studying just the \BD\ part of this extended system 
(\ref{fullBD}), we are assuming that the precursor 
chemical decays in such a way that the monomer 
concentration remains constant, that is $\dot c_1=0$.  
This requires 
\beq 
\dot p = - J_1 -  \sum_{r=1}^\infty J_r . 
\eeq 
Solutions which we describe as {\em equilibrium} solutions 
satisfy $J_r=0$ for all $r\geq1$, thus correspond to 
$p=k_b c_1/k_f$, and for this specific concentration 
of precursor chemical the additional reaction which 
adds or removes matter from the \BD\ part of the system 
will also be in equilibrium.   
Thus the equilibrium solution of our extended system is 
given by $c_r=Q_r c_1^r$ with $c_1=k_f p /k_b$, and 
at equilibrium $J_r=0$ for all $r\geq 1$. 
However for solutions we describe as {\em steady-states}, 
the fluxes $J_r=J\neq0$ where $J$ is independent ot size 
($r$), so there is a net flux of matter into the system 
(assuming $J>0$, consequently a removal of mass if $J<0$), 
and in these cases we assume that there is sufficient 
precursor chemical to sustain the steady-state.  
In this case although $\dot c_r=0$ for all $r$, 
$\dot p\neq0$ so the precursor chemical is not 
in equilibrium.  Such steady-state solutions can be 
characterised by the flux associated with them.

%-----------------------------
\section{Coarse-graining procedure} \lbl{coarse}

Following a general coarse-grained contraction 
with constant ``mesh" size $\lambda$ in aggregation number 
(so that we only retain the aggregation numbers 
$r=\Lambda_n=(n-1)\lambda+1$), 
the kinetic equations reduce to 
\beq
\dot x_n = L_{n-1} - L_n ,\;\;\; (r\geq2), \hspace{9mm}
L_n = \alpha_n x_n x_1^{\lambda} - \beta_{n+1} x_{n+1} 
,\lbl{cgBD} \eeq\beq
\alpha_n=Ta_{\Lambda_n}a_{\Lambda_n+1}\ldots 
a_{\Lambda_{n+1}-1},\hspace{10mm}
\beta_{n+1} =T b_{\Lambda_n+1} b_{\Lambda_n+2} 
\ldots b_{\Lambda_{n+1}} , \lbl{cgrates}
\eeq
where the retained coarse-grained 
cluster concentrations are relabelled as 
$x_r := c_{\Lambda_r}$ with $x_1 := c_1$ the 
monomer concentration, 
which is not involved in the coarse-graining since 
it has a special role in the \BD\ theory~\cite{cw96,rg}. The 
parameters 
$\alpha_n,\beta_n$ are the coarse-grained 
aggregation and fragmentation rates, now representing 
the addition or removal of $\lambda$ monomers to or from 
a cluster (rather than just a single monomer which 
occurs in the full \BD\ system).  This flux of matter is 
denoted by $L_n$.  In effect, the system models the
aggregation-fragmentation processes 
\beq
(\lambda+1) X_1 \rightleftharpoons X_2 , \qquad 
X_n + \lambda X_1 \rightleftharpoons X_{n+1} \quad (n\geq2) . 
\eeq
for coarse-grained clusters of size $X_{r}$. The concentration $x_n(t)$ is representative of the 
concentrations $c_r$ for 
cluster sizes $(\Lambda_{n-1}+1) \leq r \leq \Lambda_n$. 
The factor $T$ represents a change of timescale.  This factor 
was omitted in our original formulation of the 
coarse-graining procedure~\cite{cw96,wc98};
its inclusion ensures that the large time asymptotic behaviour of 
the reduced system coincides exactly with the original 
fine-grained system in the case of size-independent 
aggregation and fragmentation rates ($a_r=a$, $b_r=b$).  
The large-time asymptotics of the 
constant-monomer-concentration \BD\ system with 
constant coefficients were found by Wattis and King
using conventional asymptotic expansion methods~\cite{wk98}.   
The leading-order solution found there falls into two cases: 
for $ac_1<b$ the 
equilibrium solution $c_r=(a/b)^{r-1} c_1^r$ is approached; 
whilst for $ac_1>b$ the steady-state solution $c_r=c_1$ is 
approached.   In both cases the asymptotic solution is 
realised by a front moving forward  into larger $r$-space
(``aggregation space'') at 
speed $\dot s=|ac_1-b|$, ahead of which ($r>s(t)$) the 
concentrations are all zero, and behind which ($r<s(t)$), the 
concentrations are either at a steady-state or in 
thermodynamic  equilibrium.  In 
order for the coarse-grained contraction to preserve the 
correct speed of propagation, the timescale $T$ should take 
the value
\beq
T = \frac{(ac_1-b)}
{\lambda(a^\lambda c_1^\lambda-b^\lambda)} . 
\lbl{re-T-scale} \eeq
This temporal rescaling also ensures that the steady-state 
fluxes in the 
full and reduced systems coincide.  In the terminology of
renormalisation theory, this makes the 
coarse-graining procedure a {\em dynamical renormalisation 
transformation}.

The four properties listed in Section~\ref{BDs} above for the 
full \BD\ system 
of equations are shared by the contracted system of eqns (\ref{cgBD}):
\beq \begin{array}{rclccrcl} 
\ol{x}_n & = & Q_{\Lambda_n} x_1^{\Lambda_n} , 
&\hspace*{5mm}&
V & = & \ds\nsum x_n \left( \log \left( \ds 
\frac{x_n}{Q_{\Lambda_n}x_1^{\Lambda_n}} \right) 
- 1 \right) , \\
\ds\sum_{n=2}^\infty g_n\dot x_n & = & 
g_1L_1+\ds\nsum ( g_{n+1}-g_n ) L_n , && 
\ro & = & \ds\nsum \,[\, ( n - 1 ) \lambda + 1 \,]\, x_n , 
\end{array} \eeq
where $\alpha_n Q_{\Lambda_n} = 
\beta_{n+1} Q_{\Lambda_{n+1}}$.

Note that it is also possible to allow the mesh size to vary with 
aggregation number \cite{wc97}, that is putting 
$\lambda=\lambda_n$ transforms the system with 
constant rates ($a_r=a$, $b_r=b$) into one with non-constant 
rates (since $\alpha_r=a^{\lambda_r}$ will be a nontrivial 
function of $r$, and similarly for $\beta_{r+1}$).   For this reason 
we shall put to one side such generalised meshes in the 
present paper.

%===============================================
\section{Renormalisation structure} \lbl{renorm} 
\setcounter{equation}{0}

We now proceed to consider the renormalisation of \BD\ 
models in which the cluster rate coefficients
are of power-law form. As discussed below, this dependence is of 
immediate relevance to the description of 
surface-limited nucleation and growth processes in 
physicochemical systems.

%---------------------------
\subsection{The \BD\ system with power-law coefficients}

In the case of \BD\ systems with power-law rate coefficients, 
the rate coefficients 
for aggregation and fragmentation are respectively 
\beq
a_r=a r^p , \qquad b_{r+1}=b r^p , \lbl{prates}
\eeq 
so that the parameter $\theta=ac_1/b$ which arises in the
constant-monomer case remains useful for 
classifying dynamical behaviour.  The parameter $p$ determines 
the variability of rate with cluster size, with $p>0$ implying 
that large cluster sizes have larger aggregation and 
fragmentation rates, and $p<0$ giving rates which 
decrease with increasing cluster size. The latter case is of 
rather less physical relevance, but we still study it here for the 
sake of completeness.  Typical values for $p$ are
$p=0,\half,\rec{3},\mfrac{2}{3},1$ for the examples of 
linear chain polymerisation, coagulation kinetics in two 
space dimensions, diffusion-limited coagulation in three 
dimensions, surface-limited coagulation in three 
dimensions, and branched chain polymerisation, respectively.  
Since a cluster's volume scales with aggregation number $r$, 
if we assume that clusters are spherical then their surface 
area scales with $r^{2/3}$ and their diameter with $r^{1/3}$, 
accounting for the presence of these exponents. More general 
exponents can be manifest in other physical situations \cite{costa}.

The partition function $Q_r$ is defined by $Q_r=(a/b)^{r-1}$ 
as in the $p=0$ case considered earlier \cite{wk98}.  
The governing equations are then  
\beq
\dot c_r = J_{r-1} - J_r \;\;\; (r\geq2) , \hspace*{12mm}
J_r = r^p ( a c_1 c_r - b c_{r+1} )  , \;\;\; (r\geq1) .  \lbl{bdp}
\eeq
For $\theta\leq1$ the system approaches the steady-state
solution (which we shall also refer to as the equilibrium solution,
even though strictly speaking it is not, owing to the boundary
condition  of constant
monomer concentration imposed by the pool chemical approximation) 
given by solving $J_r=0$, that is 
\beq
c_r = \theta^{r-1} c_1 . \lbl{2equilib}
\eeq
Note that this solution is independent of $p$, although the way 
in which the steady-state solution is approached depends on $p$.

For $\theta>1$ this solution gives diverging concentrations for 
large $r$ and is hence unphysical.  
In this case the system approaches one of a family of 
time-dependent solutions to eqns~(\ref{bdp}), in which all fluxes 
are equal; $J_r=J$ for all $r$ then implies 
\beq
c_r = \theta^{r-1} c_1 \left( 1 - J \sum_{k=1}^{r-1} 
\frac{1}{b c_1 k^p \theta^k} \right) . 
\eeq
For $\theta>1$ the sum is convergent in the limit 
$r\rightarrow\infty$, and so the flux which gives 
the least singular behaviour in this limit is 
\beq
J = b c_1 \left/ \ds\ksum k^{-p} \theta^{-k} 
\right. . \lbl{2ssJ}
\eeq

The steady-state solution valid for all $r$ is 
\beq
c_r = \frac{J}{b} \sum_{k=r}^\infty k^{-p} \theta^{r-k-1} .
\lbl{2sscr} \eeq

%----------------------------------------------
\subsection{Repeated application of contraction 
procedure: renormalisation}

We have already shown how the full \BD\ system 
of equations can be reduced by a coarse-graining 
process to the subsystem of equations 
(\ref{cgBD})--(\ref{cgrates}).  We now show that these 
equations can be further reduced, by applying the 
contraction procedure to (\ref{cgBD})--(\ref{cgrates}) 
with mesh size $\mu$, defining the monomer 
concentration in the new variables by $z_1=x_1(=c_1)$.  
The new cluster concentrations $z_p$ are representative 
of the concentrations $x_n$ for $(M_{p-1}+1) \leq p \leq M_p$ 
where $\{M_p\}_{p=1}^\infty$ is the grid of retained $n$ 
values of cluster aggregation numbers, 
determined by $M_p = (p-1)\mu+1$.  Thus we find
\beq 
\dot z_p = I_{p-1} - I_p, \;\;\;(p\geq2), \hspace*{8mm}
I_p = A_p z_p z_1^{\lambda\mu} - B_{p+1} z_{p+1} , 
\lbl{rep-cont}  \eeq
where $I_p$ is the flux of material from concentration $z_p$ 
to $z_{p+1}$, and has aggregation and fragmentation rates 
determined by 
\beq
A_p = \tilde{T} \alpha_{M_p} \alpha_{M_p+1} \ldots 
\alpha_{M_{p+1}-1}  , \hspace*{9mm}   B_{p+1} = \tilde{T} 
\beta_{M_p+1} \beta_{M_p+2} \ldots \beta_{M_{p+1}} . 
\eeq
We have denoted the change of timescale by $\tilde{T}$. 
In the case of the model with constant coefficients 
($\alpha_n=\alpha$, $\beta_n=\beta$ for all $n$), in 
order for the timescales of the approach to steady-state 
or equilibrium to be correctly reproduced, the temporal 
rescaling factor $\tilde{T}$ should be chosen to be 
\beq
\tilde{T} = \frac{ (\alpha x_1^\lambda-\beta) } 
{\mu (\alpha^\mu x_1^{\lambda\mu}-\beta^\mu) } 
= \frac{(ac_1-b)}{\lambda\mu T^\mu (a^{\lambda\mu} 
c_1^{\lambda\mu} - b^{\lambda\mu})} , 
\eeq
by analogy with (\ref{re-T-scale}).  
The rate coefficients satisfy $A_p Q_{(p-1)\lambda\mu+1} = 
B_{p+1} Q_{p\lambda\mu+1}$. 
The system satisfies the four properties 
\beq \begin{array}{rclccrcl}
\ol{z}_p & = & Q_{(p-1)\lambda\mu+1} 
\ol{z}_1^{(p-1)\lambda\mu+1}
&\hspace*{5mm}& V & = & \ds \psum
z_p\left(\log\left(\ds\frac{z_p}{Q_{(p-1)\lambda\mu+1}}
\right)-1\right)\\
\ds\psum g_p\dot z_p&=&g_1 I_1 + \ds\rsum\left( 
g_{p+1}-g_p\right) I_p &&
\ro & = & \ds\psum \,\left[\, (p-1)\lambda\mu +1 
\,\right]\, z_p . 
\end{array} \eeq
So we see that the repeated contraction is identical 
to a single contraction with mesh parameter $\gamma = 
\lambda\mu$.  The system of equations (\ref{rep-cont}) 
could have been derived by a single coarse-grained 
contraction, with mesh 
\beq
z_p = c_{ _{\Gamma_r}} , \hspace{7mm} z_1=c_1 , 
\hspace{7mm} \Gamma_p = (p-1)\gamma + 1. 
\eeq
Note also that $\Gamma_p=\Lambda_{M_p}=M_{\Lambda_p}$, 
so that the coarse-graining procedure is commutative.

The combined effect on the rate coefficients and timescale 
can be verified by noting that in the case of constant 
coefficients, $A = \tilde{T} \alpha^\mu = \tilde{T} T^\mu 
a^{\lambda\mu} = \hat{T} a^{\lambda\mu}$, where 
\beq
\hat{T} = \frac{ a c_1 - b }{ \lambda\mu( a^{\lambda\mu} 
 c_1^{\lambda\mu} - b^{\lambda\mu} ) } ,  
\eeq
which is what one would expect if $\lambda$ were
replaced by $\lambda\mu$ in equation (\ref{re-T-scale}).

At the heart of renormalisation theory is the repeated 
application of a coarse-graining procedure. The analysis 
above shows that: 
(i) a contracted version of the \BD\ system can itself be 
contracted, which is crucial if we are to use ideas from 
renormalisation theory; and (ii) the effect of 
coarse-graining the 
\BD\ equations with a mesh size $\lambda$ and then 
coarse-graining the resultant equations with a mesh size 
$\mu$ is identical to a single application of the 
coarse-graining process with mesh $\gamma=\lambda\mu$.  
Thus the limit of repeated applications of the renormalisation 
coarse-graining process corresponds to the 
$\lambda\gg1$ limit of eqns~(\ref{cgBD}). This second property 
makes the ensuing analysis possible, since we do not 
have to handle systems of equations more complicated than 
the \BD\ system itself; that is, {\em the structure of the equations 
is invariant under this transformation}. 
Note that as one may have expected {\em a priori}, 
the coarse-graining does not introduce cluster--dimer or 
more general cluster--cluster interactions; rather the 
theory is generalised so that clusters interact with several 
monomers in moving from one coarse-grained set of 
cluster sizes to the next.

%---------------------------------------------
\subsection{Contraction of power law coefficients}

If the rate coefficients in the original formulation in 
eqns~(\ref{fullBD}) 
are determined by simple power laws, namely $a_r = a r^p$ 
and $b_{r+1} = b r^p$, then the coefficients in the reduced 
model are 
\beq
\alpha_n = a^\lambda \left\{ [(r-1)\lambda+1] 
[(n-1)\lambda+2] \ldots [n\lambda]  \right\}^p . 
\eeq
Thus 
\beqa
\log \alpha_n & = & \nn
\lambda \log a + p \sum_{j=1}^\lambda 
\log(n\lambda-\lambda+j) \;\; \approx \;\; 
\lambda\log a + p \int_0^\lambda 
\log(n\lambda-\lambda+x)dx \\
& = & \lambda\log a + p \lambda \,\left[\, 
\log(n\lambda) - 1 + (1\!-\!n) \log \left( 1 - 
\rec{n} \right) \right] . \lbl{rgg1app}
\eeqa
For large $n$ this asymptotes to $\log\alpha_n\sim 
\lambda \log a + p\lambda \log(\lambda n)$, 
so for simplicity we shall take $\alpha_n = 
(a \lambda^p n^p)^\lambda$, which differs slightly at small 
values of $n$.  The backward rate coefficient is then 
$\beta_{n+1} = (b\lambda^p n^p)^\lambda$.  Since  
contracting a constant density \BD\ system leaves the kinetic 
part of the equations invariant and only affects the coefficients 
in the ``constitutive equation'' for the flux, the steady-state 
solutions and large-time asymptotics 
for the contracted systems of equations can be found directly 
from the above theory.

The new system has its own $\theta$ parameter determining 
the balance between aggregation and fragmentation 
rates in the system which, for the moment, we shall call 
$\tilde{\theta} = \alpha_r x_1^\lambda 
/ \beta_{r+1} = \theta^\lambda$, so contraction of the 
system maps $\theta$ to $\theta^\lambda$.   The parameter 
$\theta$ thus plays an important role in the renormalisation 
procedure, the fixed points of this mapping corresponding to 
$\theta=0,1,\infty$; hence such systems are of 
special interest to us.  Also the contraction maps coefficients 
with exponent $p$ to those with exponent $p\lambda$. Thus 
following a contraction, there are only three limits to consider: 
small $p$ (namely $p=0$) and large $p$ (positive and negative).

\begin{figure}[hbt]
\vspace{2.5in}
\begin{picture}(300,30)(-60,0)
\thinlines
\put( 00,100){\vector(1,0){350}}
\put( 50, 00){\vector(0,1){200}}
\put(340, 80){{\Large \mbox{$\theta$}}}
\put( 35,195){{\Large \mbox{$p$}}}
\put(100,100){\circle*{10}}
\put(180,100){\circle*{10}}
\put(260,100){\circle*{10}}
\put(100, 20){\circle*{10}}
\put(180, 20){\circle*{10}}
\put(260, 20){\circle*{10}}
\put(100,180){\circle*{10}}
\put(180,180){\circle*{10}}
\put(260,180){\circle*{10}}
\thicklines
\put(180,100){\vector(1,0){40}}
\put(180,100){\vector(0,1){40}}
\put(180,100){\vector(-1,0){40}}
\put(180,100){\vector(0,-1){40}}
\put(220,100){\line(1,0){40}}
\put(140,100){\line(-1,0){40}}
\put(180,140){\line(0,1){40}}
\put(180, 60){\line(0,-1){40}}
\put(100,100){\vector(0,1){40}}
\put(100,100){\vector(0,-1){40}}
\put(100,140){\line(0,1){40}}
\put(100, 60){\line(0,-1){40}}
\put(260,100){\vector(0,1){40}}
\put(260,100){\vector(0,-1){40}}
\put(260,140){\line(0,1){40}}
\put(260, 60){\line(0,-1){40}}
\put(200,120){\vector(1,1){40}}
\put(160,120){\vector(-1,1){40}}
\put(200, 80){\vector(1,-1){40}}
\put(160, 80){\vector(-1,-1){40}}
\put(110,85){I}
\put(190,85){II}
\put(270,85){III}
\put(110,180){IV}
\put(190,180){V}
\put(270,180){VI}
\put(110,5){VII}
\put(190,5){VIII}
\put(270,5){IX}
\put(70,85){$\theta\!=\!0$}
\put(150,85){$\theta\!=\!1$}
\put(230,85){$\theta\!\gg\!1$}
\put(15,85){$p=0$}
\put(15,170){$p>0$}
\put(10,10){$p<0$}
\end{picture}
\caption{The effect of the coarse-graining contraction or, 
equivalently, the renormalisation mapping, on the parameters 
$\theta,p$ in the \BD\ model with power-law rate coefficients. 
The nine fixed points of the map are denoted I-IX, and
represent possible universality classes for the dynamics.}
\label{rggraph}
\end{figure}
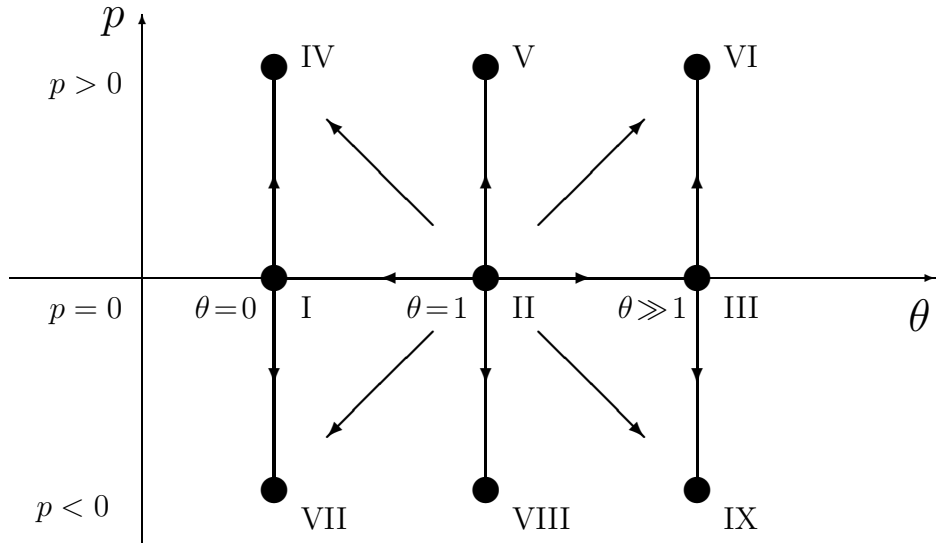

The effect of coarse-graining a \BD\ system is to modify the 
rate coefficients, by the map $\theta\mapsto 
\theta^\lambda$ and $p\mapsto\lambda p$.  If $\lambda$ 
is allowed to take on large values, there are only nine 
combinations of $(\theta,p)$ which merit attention, namely 
all possible combinations of $\theta=\{0,1,\infty\}$ and 
$p=0,p>0,p<0$.  These nine cases and their associated fixed 
points will also be the basis of analysis in subsequent 
sections of this paper.  However, the cases corresponding 
to $p<0$ are physically less relevant, since eqn 
(\ref{prates}) then assigns the largest rates to the smallest 
cluster sizes. Figure \ref{rggraph} shows schematically the 
effect of the contraction on the parameters $\theta,p$.

In phase plane terminology, II has the form of an unstable node, while 
I, III, V, VIII are saddle points, since they act as attractors 
in one direction and repellors in the other (although they are 
at the limits of the allowable domain, so only have trajectories 
on one side of the critical point). Cases IV, VI, VII, IX are stable 
nodes since they act as attractors in both the $\theta$ and 
the $p$ directions.

%------------------------------------------------
\section{Large-time asymptotics of the nine universality classes} 
\lbl{asymptotics}
\slbl{3-sec}\setcounter{equation}{0}                    % No Noise 

The previous section identified nine special, or limiting, 
cases which merit more detailed analysis, since following 
a coarse-grained contraction with large mesh parameter, 
any model will be approximated by one of these cases.  To 
simplify the analysis we shall always choose a scaling such 
that $x_1=1$, with $\alpha$ and $\beta$ equal to zero or unity.  
In this section we present large time and large aggregation-number asymptotics. These are developed by means 
of a continuum limit, which replaces the system of 
ordinary differential equations by a single partial 
differential equation, and draws on the results 
derived by King \& Wattis \cite{wk99}.  We apply these 
results in particular to our nine limiting cases. 
In a further paper these results will be generalised 
to systems whose rates are perturbed away from 
the simple polynomial expressions considered here~\cite{rgpap2}.

%--------------
\subsection{Case I: pure fragmentation at a constant rate} 
% No Noise

This corresponds to constant fragmentation with 
rate $\beta_n=1$ and no aggregation.  Hence we 
expect the system to approach the equilibrium 
configuration $x_n=0$ for all $n\geq2$.  The 
governing equation is
\beq
\dot x_n = x_{n+1} - x_n , \quad {\rm for} \;\;\; n\geq 2 .
\lbl{25Ieq} \eeq
If the system is started from the initial conditions $x_n(0)=
0$ for $n\geq2$, then the system does not change.  All other 
initial conditions will approach this state as $t\rightarrow\infty$.

%--------------
\subsection{Case II: aggregation and fragmentation 
at constant rates} 
% No Noise 

This case corresponds to constant rate coefficients but with 
both coagulation and fragmentation present, and exactly 
balancing each other, $\alpha_n=1=\beta_n$.  Thus we 
expect the system to approach the equilibrium solution 
$x_r(t)=x_1$. The time evolution follows the equation 
\beq
\dot x_n = x_{n-1} - 2 x_n + x_{n+1} , \quad {\rm for} \;\;\; n\geq 2 , 
\lbl{25IIeq} \eeq
which has the form of a ``discrete diffusion'' equation.

In the limit of large time and large $n$, the discrete system 
(\ref{25IIeq}) can be approximated by the diffusion equation 
\beq
\pad{x}{t} = \padd{x}{n} , \lbl{IIctm1}
\eeq
whose solution is $x_n=\erfc(n/2\sqrt{t})$. This equilibrium is 
approached by a purely diffusive mechanism, there being 
no advective term in (\ref{IIctm1}), in contrast to Cases I 
and III (below) where advection is present.

%--------------
\subsection{Case III: pure aggregation at a constant rate} 
% No Noise

This case corresponds to constant aggregation with rate 
coefficients $\alpha_n=1$ for all $n$, and no fragmentation 
($\beta_n=0$ for all $n$).   Thus we expect the steady-state 
solution $x_n=x_1$ to be approached, which has steady-state 
flux $L = \alpha x_n x_1^\lambda = 1$. The governing equations 
\beq
\dot x_n=  x_{n-1} - x_n  , \quad {\rm for} \;\;\; n\geq 2 , 
\lbl{25IIIeq} \eeq
are exactly solved by
\beq
x_n = 1 - e^{-t} \sum_{k=0}^{n-2} \frac{t^k}{k!} . 
\lbl{25IIIsol} \eeq
However it is not straightforward to see which mechanisms are 
driving the growth process when the solution is written in this form. An 
asymptotic approximation gives a much more intuitive idea of 
the dynamical behaviour. To describe the large time and 
large aggregation number asymptotics of this case, we 
approximate (\ref{25IIIeq}) by 
\beq
\pad{x}{t} = \half\padd{x}{n}-\pad{x}{n} , 
\eeq 
{}from which we can deduce 
the existence of a front of matter which propagates to 
larger aggregation numbers with speed unity, the front itself 
spreading out over a range of aggregation numbers 
proportional to $t^{1/2}$.  This can be shown by transforming 
to the new independent variable $z=n-t$ yielding $\pad{x}{t} 
= \half \padd{x}{z}$, which has the similarity solution 
$x=\half \erfc(z/\surd t)$. Thus in the limit $t \rightarrow\infty$ 
\beq
x_n = \half \erfc \left( \frac{n-t}{\sqrt{2t}} \right) , 
\lbl{25IIIasy} \eeq
which is an asymptotic approximation of (\ref{25IIIsol}).

%-------------==============================
\subsection{Case IV: pure fragmentation with 
rates increasing with cluster size} % No Noise

In this case the only process occurring is fragmentation, 
and the fragmentation rate depends on 
cluster size, with larger clusters losing monomers at faster 
rates.  So, as in Case I the system will tend to  $x_n=0$ 
for all $n\geq2$.  The system is governed by 
\beq
\dot x_n = n^{p\lambda} x_{n+1} - (n-1)^{p\lambda} x_n , 
\qquad n\geq2 . 
\eeq
Since the aggregation rates are greater than in Case I, 
the equilibrium solution will be approached more rapidly 
than in Case I. Matter is advected to smaller aggregation 
sizes, and removed from the system in monomeric form.

%--------------
\subsection{Case V: aggregation and fragmentation with 
rates increasing with cluster size} % No Noise

In this case we have both coagulation and fragmentation 
and both rates increase algebraically with aggregation 
number.  The system is governed by the equations 
\beq
\dot x_n = (n-1)^{p\lambda} x_{n-1} - (n-1)^{p\lambda} x_n - 
n^{p\lambda} x_n + n^{p\lambda} x_{n+1} . 
\eeq
The equilibrium solution is $x_n=1$ for all $n$, and by analogy 
with Case II (which also has $\theta=1$) we might expect 
the large-time asymptotics to approach this solution.  However, 
if we seek solutions from the wider class of steady-state 
solutions which are simply characterised by constant flux $L$ 
(not necessarily equal to zero), then we find a family of solutions 
\beq
x_n = 1 - L \sum_{k=1}^{n-1} \frac{1}{k^{p\lambda}} . 
\eeq
{}From this family, we seek the solution which has the fastest 
possibly decay as $n\rightarrow\infty$.  This occurs for $L= 
1/\zeta(p\lambda)$ (where 
$\zeta(z) = \ksum k^{-z}$ is the Riemann zeta function). 
Thus for $p\leq1/\lambda$, we 
have that $L=0$ and the equilibrium solution determines the 
large-time asymptotic behaviour of the system; however, 
for $p>1/\lambda$ there is a steady-state solution (with 
$L>0$) which has a faster decay at large $r$; hence the 
evolution will approach this solution rather than the 
equilibrium solution.

We shall now discuss the manner in which the {\em equilibrium} 
solution $x_n=1$ is approached in the case $p<1/\lambda$. 
The continuum limit for $n\gg1$ is 
\beq
\pad{x}{t} = n^{p\lambda} \left( \padd{x}{n} + 
\frac{p\lambda}{n} \pad{x}{n} \right) ,
\lbl{Vpde2}\eeq
which immediately yields a similarity solution with variable 
$\eta=n/t^{1/(2-p\lambda)}$. Assuming $x=f(\eta)$ with 
$f(0)=1$ implies 
\beq
f(\eta)=\frac{\int_\eta^\infty u^{-p\lambda} 
\exp( - u^{2-p\lambda} / (2-p\lambda)^2) du}
{\int_0^\infty u^{-p\lambda} \exp( - u^{2-p\lambda} 
/ (2-p\lambda)^2) du } . 
\lbl{Vsimsol} \eeq
In this case, there is no travelling diffusive wavefront; rather 
the balance of coagulation and fragmentation means that 
the equilibrium solution is reached by a purely diffusive 
mechanism.  However, the similarity variable $\eta$ shows 
that the aggregation number at a given concentration 
increases at a rate $n\propto t^{1/(2-p\lambda)}$, which is 
faster than the usual spreading rate of $n\propto\surd t$; 
this is due to the size-dependent coefficients accelerating 
mass transfer in the large $n$ region.   The precise details of 
the special case $p=1/\lambda$ are not covered by this 
general solution, but the special case $p=0$ (Case II) can 
be obtained from the general theory for $p<1/\lambda$.

When $p>1/\lambda$ the analysis can be modified for 
the approach to the steady-state solution 
\beq
x_n^{{\rm sss}} = \frac{1}{\zeta(p\lambda)} \sum_{k=n}^\infty 
\frac{1}{k^{p\lambda}} . \lbl{35SSS} 
\eeq
The manner in which such a solution is approached can 
be found by writing $x_n(t) = x_n^{{\rm sss}} \psi_n(t)$.  
We are then interested in the manner in which 
$\psi_n\rightarrow1$ as $t\rightarrow\infty$. 
This is governed by the equation
\beq
\dot \psi_n = \left( (n\!-\!1)^{p\lambda} + 
\frac{L}{x_n^{{\rm sss}}} \right) \psi_{n-1} - 
n^{p\lambda} \psi_n - (n\!-\!1)^{p\lambda} \psi_n + 
\left( n^{p\lambda} - \frac{L}{x_n^{{\rm sss}}}\right) \psi_{n+1} , 
\eeq
where $L$ is the steady-state flux ($L=1/\zeta(p\lambda)$).  
For large $n$ and at large times, it is appropriate to take 
the continuum limit, where $x_n^{{\rm sss}}$ can be 
approximated by $1/(p\lambda\!-\!1) n^{p\lambda-1} 
\zeta(p\lambda)$, yielding 
\beq
\frac{1}{n^{p\lambda}} \pad{\psi}{t} = \padd{\psi}{n} + 
\left(\frac{2-p\lambda}{n}\right) \pad{\psi}{n} . 
\lbl{cleanVctmsss} \eeq
This equation possesses a similarity solution of the form 
$\psi(r,t)=f(\eta)$ where $\eta=r/t^{1/(2-p\lambda)}$ and  
\beq 
f(\eta)=\ds\frac
{\ds\int_\eta^\infty u^{p\lambda-2} 
\exp (-u^{2-p\lambda}/(2\!-\!p\lambda)^2)\,du}
{ \ds \int_0^\infty  u^{p\lambda-2} 
\exp (-u^{2-p\lambda}/(2\!-\!p\lambda)^2)\,du} . 
\lbl{cleanVsimsssol} \eeq
However this solution is only defined for $1/\lambda < 
p < 2/\lambda$ and is not valid for the case $p=1/\lambda$.  
In the special case $p=2/\lambda$ equation 
(\ref{cleanVctmsss}) is solved by 
\beq
\psi = \half \erfc \left( \frac{ \log(n) - t }{2\sqrt{t}} \right) . 
\eeq
Although different states are approached in the cases 
$p<1/\lambda$ and $p>1/\lambda$, in both cases the  
asymptotics are governed by a similarity solution of the 
form $\eta=n/t^{1/(2-p\lambda)}$.

%--------------
\subsection{Case VI: pure aggregation with 
rates increasing with cluster size} % No Noise 

This case corresponds to pure aggregation, the rate 
of aggregation growing with cluster size. There is thus no 
equilibrium solution; instead the system approaches the 
steady-state solution $x_n=n^{-p\lambda}$ according to 
\beq
\dot x_n = (n-1)^{p\lambda} x_{n-1} - n^{p\lambda} x_n 
\qquad n\geq2 . \lbl{25VIeq} 
\eeq
The substitution $\psi(n,t) = n^{p\lambda} x_n$ enables 
progress to be made.  From the initial conditions $\psi_n=0$ 
for $n\geq2$, the steady-state solution $\psi_n\equiv1$ 
will be reached as time progresses.  For $p>1/\lambda$ 
the system is ill-posed due to the instantaneous transport 
of material to large aggregation numbers.  
%% - check with JRK that this statement is OK  
For $p<1/\lambda$ we assume that the equilibrium state 
is reached by a front moving from $n=1$ to large $n$ as 
time increases.  To find the form of this wave at large times, 
we take the large $n$ and large $t$ limits, replacing 
eqn (\ref{25VIeq}) by 
\beq
\pad{\psi}{t} = n^{p\lambda} \left( \half \padd{\psi}{n} - 
\pad{\psi}{n} \right) . \lbl{36pde} 
\eeq
We define the position of the front by $n=s(t)$, and 
transform to a moving coordinate frame $z=n-s(t)$. 
We now show that the two `outer' solutions $\psi=1$ 
valid for $z\rightarrow-\infty$ (that is $n/s(t)<1$) and 
$\psi=0$ which holds in the limit $z\rightarrow+\infty$ 
(that is $n/s(t)>1$)  can be joined smoothly by a transition 
layer.  Equation (\ref{36pde}) implies that $\psi(z,t)$ is 
determined by
\beq
\pad{\psi}{t} - \dot s \pad{\psi}{z} = s^{p\lambda} \left( 1 + 
\frac{z}{s} \right)^{p\lambda} \left( \half \padd{\psi}{z} - 
\pad{\psi}{z} \right) . \lbl{2VIexp} 
\eeq
The leading order terms are those involving $\pad{\psi}{z}$, 
giving $\dot s = s^{p\lambda}$, which is solved by 
\beq
s(t) \sim \left[ (1\!-\!p\lambda) t \right]^{1/(1-p\lambda)} , 
\qquad {\rm as} \;\;\; t\rightarrow\infty . \lbl{2VIssol} 
\eeq
This result can be verified by considering the density, 
$\ro=\nsum n x_n \sim \sum_1^s n^{1-p\lambda} \sim 
(1-s^{2-p\lambda})/(p\lambda-2)$, which implies that the 
density asymptotically increases according to $\dot \ro \sim 
s^{1-p\lambda} \dot s$.    However, the formula for the 
density also satisfies $\dot\ro=1+\nsum n^{p\lambda} x_n=s$. 
Thus this calculation also gives $\dot s = s^{p\lambda}$, 
confirming equation (\ref{2VIssol}).

The first correction terms in (\ref{2VIexp}) are 
\beq
\frac{1}{s^{p\lambda}}\pad{\psi}{t} = 
\frac{1}{2} \padd{\psi}{z} - 
\frac{p\lambda z}{s}\pad{\psi}{z} . \lbl{VIpde2} 
\eeq
Since $s(t)$ is monotonically increasing, we simplify further 
calculations by using $s$ as our temporal variable. 
The next stage of the calculation is straightforward if 
$p<1/2\lambda$, since in this case $z\sim \sqrt{s}$ 
and (\ref{VIpde2}) reduces to $\pad{\psi}{s} = \half 
\padd{\psi}{z}$.  Thus we find 
\beq
x_n = \half n^{-p\lambda} \erfc \left( 
\frac{(n-s) \sqrt{1-2p\lambda}}{\sqrt{2s}} \right) . 
\eeq
For $p>1/2\lambda$, advection of the front dominates 
the diffusion processes and so the front retains its memory 
of the initial conditions.  An alternative way of visualising 
this process is that the front moves so fast that it reaches 
$n=\infty$ before it has had time to adjust its shape to that 
of the erfc function.  Thus we cannot specify in detail the 
shape of the front. More details of the special case 
$p=1/2\lambda$ can be found in \cite{wk99}.

%--------------===============================
\subsection{Case VII: pure fragmentation with 
rates decreasing with cluster size} % No Noise 

Here there is no aggregation, so fragmentation will cause all 
clusters to break up, and as in Cases I and IV, the system will 
approach the state where all matter is in monomeric form 
($x_n=0$ for $n\geq2$).  Since the fragmentation decays 
rapidly with size, large clusters will take a very long time to 
break up. The determining equations are 
\beq
\dot x_n = n^{p\lambda} x_{n+1} - (n-1)^{p\lambda} x_n , 
\qquad n\geq2 . 
\eeq
Since fragmentation is weaker than in Cases I 
and IV, Case VII takes longer to reach equilibrium.

%--------------
\subsection{Case VIII: aggregation and fragmentation with 
rates decreasing with cluster size} % No Noise 

This corresponds to coagulation and fragmentation being 
equally balanced, both decaying as the aggregation 
number increases. As in Cases II and V (with $p<1/\lambda$), 
the system approaches the equilibrium solution $x_n\equiv1$ 
in the large-time limit.  The similarity solution valid for 
$p<1/\lambda$ derived in Case V is valid for this case too.  
The balance of coagulation and fragmentation forces 
means that the equilibrium solution is reached by a 
predominantly diffusive process.

%----------------------------
\subsection{Case IX: pure aggregation with 
rates decreasing with cluster size} % No Noise 

The final case has no fragmentation, only an aggregation 
term which decays with increasing particle size. The system 
thus has no equilibrium solution, and will evolve to the 
divergent steady-state solution $x_n = n^{-p\lambda}$ 
according to 
\beq
\dot x_n = (n-1)^{p\lambda} x_{n-1} - n^{p\lambda} x_n , 
\qquad n\geq 2 . 
\lbl{2IXodes} \eeq
The substitution $\psi_n=n^{p\lambda} x_n$ yields 
$n^{-p\lambda}\dot\psi_n = \psi_{n-1}-\psi_n$, and 
progress can be made by forming a continuum 
approximation 
\beq
n^{-p\lambda} \pad{\psi}{t} = \half \padd{\psi}{n} - 
\pad{\psi}{n} . 
\lbl{2IXeq2} \eeq
As in previous pure aggregation cases, the large-time 
asymptotic approach to the steady-state is governed 
by a diffusive wave moving in the $n$-domain from 
$n=1$ to large $n$; also as $t$ increases the wavefront 
widens.  The speed of propagation can be found by 
considering the density injected into the system, or by 
substituting $n=s(t)+z$ into equation (\ref{2IXeq2}) and expanding 
in $z$. The leading order terms are 
\beq
\frac{1}{s(t)^{p\lambda}} \pad{\psi}{t} = 
\half \padd{\psi}{z} + \pad{\psi}{z} \left( 
s(t) ^{-p\lambda} \dot s(t) - 1 - \frac{p\lambda z}{s} \right) .  
\lbl{2IXpde} \eeq
At leading-order the dominant terms are those involving 
$\pad{\psi}{z}$, giving $\dot s = s^{p\lambda}$ since we 
assume $z/s\ll1$.  The position of the front is given by 
\beq
s(t) \simeq \left[ (1-p\lambda) t \right]^{1/(1-p\lambda)} , 
\qquad {\rm as} \;\;\; t\rightarrow \infty , 
\lbl{39ssol} \eeq
and we now use $s$ as a new timescale to solve the next 
order terms in (\ref{2IXpde}), which yield $\psi=\half\erfc( z 
\surd((1-2p\lambda)/(2s)))$. Thus 
\beq
x_n(t) \sim \half n^{-p\lambda} \erfc \left( \frac{ [n-s(t)] 
\sqrt{1-2p\lambda}}{\sqrt{2 s(t)}} \right) , \qquad {\rm as} 
\;\;\; t\rightarrow\infty \;\;\; {\rm with} \;\;\; n-s(t) = 
{\cal O}\left(\sqrt{t}\right) . \lbl{39xnsol} 
\eeq
{}From equation (\ref{39ssol}) it is seen that the effect of $\lambda$ 
is to reduce the exponent with which the position of the 
wavefront scales with time.  Equation (\ref{39xnsol}) 
shows that $\lambda$ increases the decay exponent 
of the steady-state solution from $p$ to $p\lambda$,
and alters the scaling of the width of the wave.  In the fully 
detailed system of equations the width asymptotes to 
$\sqrt{2s/(1-2p)}$, which, under coarse-graining, one might
expect to be mapped to $\sqrt{2s/(1-2p)}/\lambda$; 
however, solving the coarse-grained equations yields 
$\sqrt{2s/(1-2p\lambda)}$ instead.

%----------------------------------------
\section{Discussion} \lbl{discussion}
\setcounter{equation}{0}

We have described in detail the nine archetypal classes
of behaviour into which the asymptotic dynamics of the
\BD\ equations with power-law coefficients falls. These 
nine classes  (``Cases I--IX'' of Section~\ref{asymptotics})
arise as a result of the application of  
a systematic renormalisation procedure 
to the fine-grained \BD\ equations and capture 
qualitatively different physical properties which are 
shared by all the models within the class, regardless 
of their microscopic differences. This amounts
to a massive simplification of the original problem. 
In the present paper, we have concentrated on 
the detailed large-time asymptotic analysis of these 
nine cases.

One specific detailed feature is worthy of note here. Our analysis of 
Cases IV, V and VI shows that coarse-graining 
retains the correct leading-order structure of the problem, 
though critical exponents of $p=1$ where the 
behaviour changes from fragmentation to aggregation dominated 
are mapped to $p=1/\lambda$. 
Thus if $p<1$, to retain the correct qualitative behaviour 
one is limited to taking $\lambda<1/p$: taking the 
limit $\lambda\rightarrow\infty$ causes the 
correct form of the kinetics to be lost, although the correct solution 
is still be approached in Cases IV and VI. 
In Case V, the equilibrium solution $x_n=1$ is approached if 
$\lambda<1/p$, and a steady-state solution if $p>1$.  
However, if $\lambda>1/p>1$, the coarse-grained system 
approaches a steady-state solution whereas the microscopic 
model approaches the equilibrium solution.

In a companion paper~\cite{rgpap2} we analyse the 
effect that random perturbations to the rate coefficients 
have on the above analysis.  There we consider the modifications 
such perturbations make to the
steady-state solutions, the kinetics of approach to these 
states and their effects on coarse-graining process itself. That
analysis is necessary to confirm the existence and number of distinct
universality classes in these \BD\ systems; as we have previously
announced, all nine classes of behaviour may indeed be described as
such (modulo, in Cases VII-IX, a particular requirement on the
behaviour of the noise)~\cite{rg}.

The renormalisation programme described here is of direct relevance to
many of our recent publications on the application of generalised \BD\
models to a wide range of problems of physicochemical interest. A
further paper currently in preparation~\cite{expo} explores
several of these nucleation and growth problems in which the 
renormalisation scheme is {\em exact}.

%-----------------------------------------------
\subsection*{Acknowledgements}

We are grateful to John Cardy for several useful discussions 
regarding renormalisation theory, and to Bob O'Malley for 
pointing out related work in the literature.   PVC is grateful 
to the Department of Theoretical Physics and Wolfson College, 
University of Oxford, for a Wolfson-Harwell Visiting Fellowship, 
and for hosting our discussions.  JADW thanks John King 
for many instructive conversations.

%------------------------------------------------

\normalsize

%------------------------------------------------
%\newpage 

\listoffigures

\begin{figure}[hbt]
\vspace{3in}
\begin{picture}(300,30)(-60,0)
\thinlines
\put( 00,100){\vector(1,0){350}}
\put( 50, 00){\vector(0,1){200}}
\put(340, 80){{\Large \mbox{$\theta$}}}
\put( 35,195){{\Large \mbox{$p$}}}
\put(100,100){\circle*{10}}
\put(180,100){\circle*{10}}
\put(260,100){\circle*{10}}
\put(100, 20){\circle*{10}}
\put(180, 20){\circle*{10}}
\put(260, 20){\circle*{10}}
\put(100,180){\circle*{10}}
\put(180,180){\circle*{10}}
\put(260,180){\circle*{10}}
\thicklines
\put(180,100){\vector(1,0){40}}
\put(180,100){\vector(0,1){40}}
\put(180,100){\vector(-1,0){40}}
\put(180,100){\vector(0,-1){40}}
\put(220,100){\line(1,0){40}}
\put(140,100){\line(-1,0){40}}
\put(180,140){\line(0,1){40}}
\put(180, 60){\line(0,-1){40}}
\put(100,100){\vector(0,1){40}}
\put(100,100){\vector(0,-1){40}}
\put(100,140){\line(0,1){40}}
\put(100, 60){\line(0,-1){40}}
\put(260,100){\vector(0,1){40}}
\put(260,100){\vector(0,-1){40}}
\put(260,140){\line(0,1){40}}
\put(260, 60){\line(0,-1){40}}
\put(200,120){\vector(1,1){40}}
\put(160,120){\vector(-1,1){40}}
\put(200, 80){\vector(1,-1){40}}
\put(160, 80){\vector(-1,-1){40}}
\put(110,85){I}
\put(190,85){II}
\put(270,85){III}
\put(110,180){IV}
\put(190,180){V}
\put(270,180){VI}
\put(110,5){VII}
\put(190,5){VIII}
\put(270,5){IX}
\put(70,85){$\theta\!=\!0$}
\put(150,85){$\theta\!=\!1$}
\put(230,85){$\theta\!\gg\!1$}
\put(15,85){$p=0$}
\put(15,170){$p>0$}
\put(10,10){$p<0$}
\end{picture}

\end{figure}


\begin{thebibliography}{99}
%
\addcontentsline{toc}{section}{References}
\renewcommand{\itemsep}{-1pt}
\renewcommand{\parsep}{-1pt}
%
\bibitem{amit} DJ Amit. {\em Field theory, the renormalization 
group and critical phenomena}. McGraw-Hill, London, (1978).
%
\bibitem{b+m} JA Blackman \& A Marshall. Coagulation 
and fragmentation in cluster monomer reaction models. 
{\em J Phys A Math Gen}, {\bf 27}, 725--740, (1994).  
%
\bibitem{bk91} J Bricmont \& A Kupiainen. 
Renormalisation group for diffusion in a random medium. 
{\em Phys. Rev. Lett.}, {\bf 66}, 1689-1692, (1991).
%
\bibitem{bk92} J Bricmont \& A Kupiainen. 
Renormalisation group and the Ginzburg-Landau equation. 
{\em Commun. Math. Phys.}, {\bf 150}, 193-208, (1992).
%
\bibitem{bk94} J Bricmont, A Kupiainen \& G Lin. 
Renormalisation group and asymptotics of solutions 
of nonlinear parabolic equations. {\em Commun. on 
Pure \& Applied Mathematics}, {\bf 47}, 893-922, (1994).
%
\bibitem{bk} NV Brilliantov \& PL Krapivsky. 
Non-scaling and source-induced scaling behaviour 
in aggregation models of movable monomers and 
immovable clusters. {\em J. Phys. A.}, {\bf 24}, 
4787--4803, (1991).
%
\bibitem{cardy} J Cardy.  {\em Scaling and Renormalization 
in Statistical Physics.} Cambridge Lecture Notes in Physics, 
Vol.~5, CUP, (1996).
%
\bibitem{carr} J Carr. Asymptotic behaviour of solutions to the 
coagulation-fragmentation equations I, the strong fragmentation 
case. {\em Proc Roy Soc Ed}, {\bf 121A}, 231-244, (1992). 
%
\bibitem{cc} J Carr \& FP da Costa. Asymptotic behaviour of 
solutions to the coagulation-fragmentation equations II, the 
weak fragmentation case. {\em J Stat Phys}, {\bf 77}, 89-123, (1994).
%
\bibitem{fpc} FP da Costa.  Existence and 
uniqueness of density-conserving solutions to the 
coagulation-fragmentation equations with strong fragmentation. 
{\em J Math Anal \& Appl}, {\bf 192}, 892-914, (1995). 
%
\bibitem{costa} FP da Costa.  A finite dimensional dynamical 
model for gelation in coagulation processes.
{\em J. Nonlinear Sci.}, {\bf 8}, 619--653, (1998). 
%
\bibitem{ch90} PV Coveney  \& RR Highfield.  {\em 
The Arrow of Time}. W.H.~Allen, London, (1990).
%
\bibitem{ch95} PV Coveney  \& RR Highfield.  {\em 
Frontiers of Complexity}. Faber and Faber, London; 
Fawcett, New York, (1995).
%
\bibitem{cw96} PV Coveney \& JAD Wattis.   
Analysis of a generalised \BD\ model of 
self-reproducing micelles.  {\em Proc. Roy.
Soc. Lond. A}, {\bf 452}, 2079--2102, (1996).
%
\bibitem{cw98} PV Coveney \& JAD Wattis. 
A Becker-D\"{o}ring model of self-reproducing 
vesicles.  {\em J. Chem. Soc.: Faraday Trans.}, 
{\bf 102}, 233--246, (1998)
%
\bibitem{rg} PV Coveney \& JAD Wattis. Cluster 
renormalisation for the \BD\ equations. {\em J. 
Phys. A: Math. Gen.}, {\bf 32}, 7145--7152, (1999). 
%
\bibitem{dkw98} SC Davies, JR King \& JAD Wattis. 
Self-similar Behaviour in the Coagulation Equations. 
{\em J. Eng. Math.}, {\bf 36}, 57--88, (1999). 
%
\bibitem{dongen+e} PGJ van Dongen \& MH Ernst.  
Dynamic scaling in the kinetics of clustering. 
{\em Phys Rev Lett} {\bf 54}, 1396--1399, (1985). 
%
\bibitem{dongen+e2} PGJ van Dongen \& MH Ernst.  
Cluster size distribution in irreversible aggregation at large 
times. {\em J Phys A Math Gen}, {\bf 18}, 2779--2793, (1985).  
%
\bibitem{dongen+e3} PGJ van Dongen \& MH Ernst.  
Kinetics of reversible polymerization. 
{\em J Stat Phys}, {\bf 37} 301--324, (1984). 
%
\bibitem{goldbook} N Goldenfeld. Lectures on Phase 
Transitions and the Renormalization Group. Frontiers in 
Physics, 85. Addison Wesley, (1992).
%
\bibitem{hend} EM Hendriks \& MH Ernst.  
Exactly soluble addition and condensation models in 
coagulation kinetics. {\em Journal of Colloid and 
Interface Science}, {\bf 97}, 176--194, (1984)
%
\bibitem{itzyk80} C Itzykson \& J-B Zuber. {\em Quantum 
Field Theory}. McGraw-Hill Inc., New York (1980).
%
\bibitem{krap} PL Krapivsky.  Kinetics 
of catalytic reactions with diffusional relaxation. 
{\em Phys Rev E}, {\bf 52A}, 3455--3461, (1995). 
%
\bibitem{krap+r} PL Krapivsky \& S Redner.  Transitional 
aggregation kinetics in dry and damp environments. 
{\em Phys Rev E}, {\bf 54A}, 3553--3561, (1996). 
%
\bibitem{peliti} L Peliti.  Renormalization of fluctuation 
effects in the $A+A\rightarrow A$ reaction. 
{\em J Phys A Math Gen}, {\bf 19}, L365--L367, (1986). 
%
\bibitem{pl76} Penrose, O. \& Lebowitz, J.L. \ \ Towards a 
Rigorous Theory of Metastability, in {\em Studies in Statistical 
Mechanics VII: Fluctuation Phenomena}, pp.322--375. Montroll, 
E. \& Lebowitz, J.L. (eds), North Holland, Amsterdam, (1976). 
%
\bibitem{sp} RW Samsel \& AS Perelson. Kinetics of rouleau 
formation. I. A mass action approach with geometric 
features. {\em Biophys J}, {\bf 37}, 493--514, (1982). 
%
\bibitem{velaz} JJL Velazquez. The \BD\ equations 
and the Lifshitz-Slyozov theory of coarsening. 
{\em J. Stat. Phys.}, {\bf 92}, 195--236, (1998). 
%
\bibitem{compBD} JAD Wattis. A Becker-D\"{o}ring 
model of competitive nucleation.  {\em J. Phys A:
Math Gen.}, {\bf 32}, 8755-8784, (1999). 
%
\bibitem{wc97} JAD Wattis \& PV Coveney.  General 
nucleation theory  with inhibition for chemically 
reacting systems. {\em J. Chem. Phys.}, {\bf 106}, 
9122-9140, (1997).
%
\bibitem{wc98} JAD Wattis \& PV Coveney.  The 
origin of the RNA world: a kinetic model. {\em 
J. Phys. Chem. B}, {\bf 103}, 4231--4250, (1999).
%
\bibitem{wc99} JAD Wattis \& PV Coveney. Models of nucleation 
and growth mechanisms: the challenge to experiment. 
{\em Phys. Chem., Chem. Phys.}, {\bf 1}, 2163--2176, (1999).
%
\bibitem{rgpap2} JAD Wattis \& PV Coveney. 
Renormalisation-theoretic analysis of non-equilibrium 
phase transitions II: the effect of perturbations on rate 
coefficients in the \BD\ equations. {\em submitted to 
J Phys A}, (2001).
%
\bibitem{expo} JAD Wattis \& PV Coveney. 
Exact renormalisation theory for the \BD\ equations with
size-dependent rate coefficients. 
{\em In preparation}, (2001). 
%
\bibitem{wk98} JAD Wattis \& JR King. 
Asymptotic solutions to the \BD\ equations. 
{\em J. Phys. A.}, {\bf 31}, 7169--7189, (1998).
%
\bibitem{wk99} JAD Wattis \& JR King.  Asymptotic 
solutions to the \BD\ equations with non-constant 
coefficients.  {\em In preparation}, (2001). 
%
\bibitem{wilson} KG Wilson. {\em Phys. Rev. B}, 
{\bf 4}, 3174 \& 3184 (1971); KG Wilson \& J Kogut, 
{\em Phys. Rep. C}, {\bf 12}, 75 (1974).
%
\bibitem{w93} SL Woodruff. The use of an invariance 
condition in the solution of multiple-scale singular 
perturbation problems: ordinary differential equations. {\em 
Studies in Applied Mathematics}, {\bf 90}, 225-248, (1993). 
%
\bibitem{w95} SL Woodruff. A uniformly valid 
asymptotic solution to a matrix system of ordinary 
differential equations and a proof of its validity. {\em 
Studies in Applied Mathematics}, {\bf 94}, 393-413, (1995). 
%
\bibitem{w95b} SL Woodruff. Multiple-scale 
perturbation analysis of the direct interaction 
approximation for inertial-range turbulence. 
{\em Phys Rev E}, {\bf 52}, 3245-3248, (1995). 
%
\bibitem{wm94} SL Woodruff \& AF Messiter. 
A perturbation analysis of an interaction between 
long and short surface waves. {\em Studies in 
Applied Mathematics}, {\bf 92}, 159-189, (1994). 
%
\end{thebibliography}
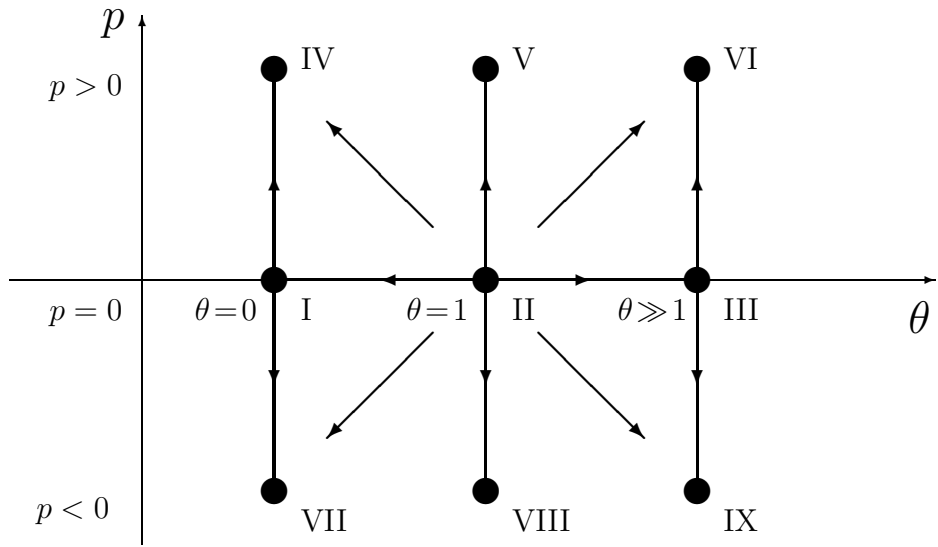
\end{document}